\pdfoutput=1
\documentclass[prd,superscriptaddress,nofootinbib]{revtex4}
\usepackage{graphicx}
\usepackage{epsfig}
\usepackage{bm}
\usepackage{hyperref}
\usepackage{latexsym,amssymb,amsmath,amsfonts,amssymb,txfonts,pxfonts,wasysym,float}
\usepackage{mathrsfs}
\usepackage{color}
\usepackage{lettrine}
\usepackage{lipsum}
\usepackage{enumitem}
\usepackage{verbatim}

 \def\be{\begin{equation}}
\def\ee{\end{equation}}
\def\ba{\begin{eqnarray}}
\def\ea{\end{eqnarray}}
\usepackage{pdfsync}

\usepackage[usenames,dvipsnames]{xcolor}
\definecolor{orange}{cmyk}{0,0.5,1,0}
\definecolor{rossoCP3}{cmyk}{0,.88,.77,.40}
\definecolor{graa}{rgb}{0.8,0.8,0.8}
\definecolor{blaa}{rgb}{0.2,0.2,0.6}

\def\simlt{\mathrel{\lower2.5pt\vbox{\lineskip=0pt\baselineskip=0pt
           \hbox{$<$}\hbox{$\sim$}}}}
\def\simgt{\mathrel{\lower2.5pt\vbox{\lineskip=0pt\baselineskip=0pt
           \hbox{$>$}\hbox{$\sim$}}}}

\begin{document}

\title{\color{rossoCP3} Using Neutrino Oscillations to Measure $\bm{H_0}$?}

\author{Luis A. Anchordoqui}

\affiliation{Department of Physics and Astronomy,  Lehman College, City University of
  New York, NY 10468, USA}

\affiliation{Department of Physics,
 Graduate Center, City University
  of New York,  NY 10016, USA}

\begin{abstract}
  \vskip 2mm \noindent Recently, 
  the idea of
  using neutrino oscillations to measure the Hubble constant was introduced. 
   We show that such a
  task is unfeasible because for typical energies of cosmic
neutrinos, oscillations average out over cosmological distances and so
the oscillation probability depends only on the mixing angles.
\end{abstract}

\maketitle

\section{Introduction}

In the course of the last decade, an intriguing inconsistency between measurements of
the cosmic expansion rate based on early- and late-Universe probes
has emerged. This inconsistency shows up as a discrepancy in the
value of the Hubble constant $H_0$~\cite{Hubble:1929ig}, as inferred from measurements of
the anisotropy of the cosmic microwave background, $H_0=\left(67.27\pm
  0.60\right){\rm \,km\,s^{-1}\,Mpc^{-1}}$ at 68\%
CL~\cite{Planck:2018vyg}, and as measured from a series of distance
indicators in the local Universe. More precisely, the latest distance
ladder measurement based on Type Ia supernovae (SNIa) calibrated by
Cepheids gives $H_0=(73.04 \pm 1.04){\rm \,km\,s^{-1}\,Mpc^{-1}}$ at
68\% CL~\cite{Riess:2021jrx}, whereas using the Tip of the Red Giant
Branch to calibrate SNIa leads to $H_0=(72.4\pm2.0)~{\rm km \,s^{-1}\,
  Mpc^{-1}}$~\cite{Yuan:2019npk} and  $H_0=(69.6 \pm 0.8\,({\rm
stat}) \pm 1.7\,({\rm sys})){\rm  km\,s^{-1}\, Mpc^{-1}}$~\cite{Freedman:2020dne}, both at 68\%~CL. Depending on which set of measurements one combines,
the tension between the model-dependent and independent estimates of $H_0$ sits between $4.5\sigma$ to $6.3\sigma$~\cite{Abdalla:2022yfr}. This so-called ``$H_0$ tension'' has become the new cornerstone of modern cosmology,
and many new-physics setups are rising to the challenge~\cite{DiValentino:2021izs}.  

In a recent publication,
 the possibility of using
neutrino oscillations as a distance indicator to measure $H_0$~\cite{Khalifeh:2021jfl} was entertained.
 In~this article,
  we show that such a task is unfeasible because for typical energies of neutrinos originating via cosmic-ray interactions in astrophysical sources, oscillations average out over  cosmological distances and so
the oscillation probability depends only on the angles of the leptonic mixing~matrix. 

The layout of the paper is as follows. 
In~Section~\ref{sec:2}, we go through the formalism of neutrino oscillations and show that oscillation averaged probabilities for transition between flavors have no dependence on the travel distance to the astrophysical sources of cosmic neutrinos. In~Section~\ref{sec:3}, we examine the conditions for the coherence loss of the neutrino wavepacket. Finally, in~Section~\ref{sec:4}, we present our~conclusions.

\section{Oscillations of High-Energy Cosmic~Neutrinos} \label{sec:2}

Neutrino oscillations are the outcome of nonzero neutrino masses and the certainty that virtually all useful neutrino sources are coherent. In~other words, the~neutrinos produced via charged-current weak interactions associated with $l_\alpha$ charged leptons, $\alpha = e, \mu, \tau$, can be described as coherent superpositions of neutrino states $\nu_j$ with different masses $m_j,\, j = 1, 2, 3$, weighted by the elements $U_{\alpha j}$ of the Pontecorvo--Maki--Nagakawa--Sakata (PMNS)
matrix~\cite{Pontecorvo:1967fh,Pontecorvo:1957qd,Maki:1962mu}, i.e.,
\begin{equation}
|\nu_\alpha \rangle = \sum_j U^*_{\alpha j} \ |\nu_j\rangle \, .
\end{equation}
The superposition of mass eigenstates is valid in no small part because  neutrino masses are tiny when compared with their laboratory/cosmic energies. For~a neutrino of energy $E_\nu$ traveling a distance $L$, we can conveniently parameterize the oscillation phase $\Delta_{ij}$ as
\begin{equation}
\Delta_{ij} = \frac{\Delta m_{ij} \ L}{4 E_\nu} \simeq 1.27 \
\Bigg(\frac{\Delta m^2_{ij}}{{\rm eV}^2}\Bigg) \ \Bigg(\frac{L}{{\rm
    km}}\Bigg) \ \Bigg(\frac{E_\nu}{{\rm GeV}}\Bigg)^{-1} \,,
\label{Deltaij}
\end{equation}
where $\Delta m^2_{ij} \equiv m_i^2 -
m_j^2$~\cite{Anchordoqui:2013dnh}. Now, for~a relatively low redshift of $z=0.05$,
the distance traveled by the neutrinos is about 200~Mpc. Taking the
highest energy measured at the IceCube facility $E_\nu \sim 10^7~{\rm
  GeV}$~\cite{IceCube:2020wum} and the solar mass splitting, $\Delta m^2_{ij} \equiv \Delta
m_\odot^2 \simeq 7.42 \times 10^{-5}~{\rm eV}^2$~\cite{Gonzalez-Garcia:2021dve}, Eq.~(\ref{Deltaij})
leads to $\Delta_{ij} \sim 6 \times 10^{10}$, which is the number of
oscillation periods the neutrinos would experience. That is, the~neutrinos would experience 60 billion oscillations to the Earth over a
redshift $z=0.05$. This implies that we would need to know $\Delta
m_\odot^2$ to one part in 60 billion and we would also have to measure
$E_\nu$ to one part in 60~billion. Both of these measurements are
ridiculously unfeasible. 
After~averaging over these uncertainties,
 we 
find the so-called oscillation averaged probability for transition 
of flavor $\alpha$ to $\beta$, which is given by
\begin{equation}
P(\nu_\alpha \to \nu_\beta) = \sum_{i} U_{\alpha i}^2 U_{\beta i}^2 \,,
\label{Palphabeta}
\end{equation}
and has no dependence on $L$~\cite{Anchordoqui:2013dnh}. Note that for
$z = 0.05$, an~oscillation phase $\Delta_{ij} \sim 1$ would require an
unrealistic neutrino energy of roughly $10^{16}~{\rm GeV}$.

\section{Coherence Loss of Cosmic Neutrino~Oscillations}
\label{sec:3}

There is also a loss of coherence  during neutrino propagation~\cite{Kiers:1995zj,Stodolsky:1998tc}. This is because different neutrino mass eigenstates of the same energy
have different velocities, and~so the wavepackets of the
mass eigenstates composing a neutrino state will come apart as they
propagate. For~neutrinos traveling over cosmological distances, the~flying  path is so large that these components completely separate from each other. In~the case of coherence loss, the~oscillatory terms of the oscillation probabilities~disappear. 

Following~\cite{Farzan:2008eg},
 we define $\Delta E_{\nu,L}$ as the energy difference for which
\begin{equation}
  \Delta_{ij} (E_\nu-\Delta E_{\nu,L},L) - \Delta_{ij} (E_\nu,L) = \pi \, .
\end{equation}
A straightforward calculation leads to
\begin{equation}
  \Delta E_{\nu,L} \simeq \frac{4 \pi E_\nu^2}{\Delta m_{ij}^2 L} = E_\nu \frac{\ell_{ij}}{L} \,,
\end{equation}
where $\ell_{ij} = 4 \pi E_\nu/\Delta m_{ij}^2$ is the vacuum oscillation length. As~neutrinos travel cosmological distances between their origins and us,
 they are essentially on their mass shells, satisfying $E_\nu^2 = p^2 + m_j^2$. The~ relativistic dispersion relation implies that $E_\nu \sigma_{E_\nu} = p \sigma_p$, where $\sigma_{E_\nu}$ and $\sigma_p$ are the uncertainties in the energy and the width of the wavepacket in the longitudinal direction. Because~neutrinos are ultrarelativistic $E_\nu \simeq p$, and~hence 
$\sigma_{E_\nu} \simeq \sigma_p$. It is easily seen that the interference between the effects of different mass eigenstates disappears if
\begin{equation}
  \sigma_p > \Delta E_{\nu,L} = E_\nu \frac{\ell_{ij}}{L} \, ;
\end{equation}
an
 equivalent expression can be found in configuration space, where the size of the wavepacket at the production point is given by the inverse of the uncertainty $\sigma_p$, i.e.,~$\sigma_x \sim \sigma_p^{-1}$~\cite{Farzan:2008eg}.

If the parent pion does not undergo any interaction with matter or with the magnetic field before decay, the~width of the wavepacket in the configuration space is estimated to be
\begin{equation}
  \sigma_x  \sim 2 \times 10^{-6} \left(\frac{E_\nu}{10^7~{\rm GeV}}\right)^{-1}~{\rm cm}
\end{equation}
whereas for a parent muon,
\begin{equation}
  \sigma_x \sim 2 \times 10^{-4} \left(\frac{E_\nu}{10^7~{\rm GeV}}\right)^{-1}~{\rm cm} \,,
\end{equation}
i.e., for~the neutrinos produced by muons, $\sigma_x$ is larger by the ratio of the lifetime of the muon to that of the pion~\cite{Farzan:2008eg}. The~dominant modification of $\sigma_x$ at the sources comes from the interaction of the parent charged particle with the magnetic field $B$, which can be parameterized as
\begin{equation}
  \sigma_x \sim 6 \times 10^{-19} \left(\frac{\Gamma}{100}\right)^{1/2} \left(\frac{B}{10^7~{\rm GeV}}\right)^{-1/2} \left(\frac{E_\nu}{10^7~{\rm GeV}}\right)^{-3/2}~{\rm cm} \,,
    \end{equation}
for pions and
\begin{equation}
  \sigma_x \sim 5 \times 10^{-17} \left(\frac{\Gamma}{100}\right) \ \left(\frac{B}{10^7~{\rm G}}\right)^{-1} \, \left(\frac{E_\nu}{10^7~{\rm GeV}}\right)^{-2}~{\rm cm} \,,
\end{equation}
for muons, where $\Gamma$ is the Lorentz boost of the plasma~\cite{Farzan:2008eg}.

A straightforward calculation shows that for typical $B$ fields of cosmic neutrino sources, the~mass states will decohere after propagation over cosmological distances, i.e.,~the mass eigenstates will be separated enough that there is no overlap of the wavefunctions and we detect on Earth the three mass states each with their own probability
given by Eq.~(\ref{Palphabeta}). Indeed, as~shown in~\cite{Ohlsson:2000mj},
  oscillation averaging
and full decoherence yield the same probability. This implies that it
does not even matter if we measure $E_\nu$ and $\Delta m_{ij}^2$ to
arbitrary~precision.

\section{Conclusions} \label{sec:4}

We have shown that for typical energies of cosmic
neutrinos, oscillations average out over cosmological distances and so
the oscillation probability depends only on the mixing angles of the PMNS matrix. 
As~a consequence, neutrino
oscillations cannot be used to estimate cosmological distances and therefore cannot be adopted as a probe to measure $H_0$.

\acknowledgments{Work supported by the 
  U.S. National Science Foundation (NSF Grant PHY-2112527).}

\end{document}